\documentclass[onecolumn]{revtex4}

\input{epsf}

\def\AJ{{\it Ap. J.} }

\def\CQG{{\it Class. Quantum Gravity} }

\def\MPL{{\it Mod. Phys. Lett.} }

\def\PR{{\it Phys. Rev.} }
\def\PRL{{\it Phys. Rev. Lett.} }

\def\al{\alpha} \def\be{\beta} \def\ga{\gamma} 
\def\ep{\epsilon}   
\def\th{\theta}   
\def\la{\lambda}

\def\om{\omega}  \def\De{\Delta} 
\def\La{\Lambda}   
 \def\Om{\Omega}

 \def\frac#1#2{{\textstyle{{#1}\over
{#2}}}} 
\def\lsim{\mathrel{\rlap{\lower4pt\hbox{\hskip1pt$\sim$}}
\raise1pt\hbox{$<$}}}
\def\gsim{\mathrel{\rlap{\lower4pt\hbox{\hskip1pt$\sim$}}
\raise1pt\hbox{$>$}}} \def\sqr#1#2{{\vcenter{\vbox{\hrule height.#2pt
\hbox{\vrule width.#2pt height#1pt \kern#1pt \vrule width.#2pt} \hrule
height.#2pt}}}}

\def\beq{\begin{equation}} \def\eeq{\end{equation}}
\def\beqa{\begin{eqnarray}} \def\eeqa{\end{eqnarray}}

\long\def\symbolfootnote[#1]#2{\begingroup%
\def\thefootnote{\fnsymbol{footnote}}\footnote[#1]{#2}\endgroup}

\begin{document}

\title{Galileo satellite constellation and extensions to General Relativity}

\vskip 0.2cm

\author{J. P\'aramos\symbolfootnote[4]{Also at Centro de F\'isica de Plasmas, Instituto
Superior T\'ecnico.}} \email{x_jorge@fisica.ist.utl.pt}

\author{O. Bertolami\footnotemark[4]} \email{orfeu@cosmos.ist.utl.pt}

\vskip 0.2cm

\affiliation{Instituto Superior T\'ecnico, Departamento de
F\'{\i}sica, \\Av. Rovisco Pais 1, 1049-001 Lisboa, Portugal
}

\vskip 0.2cm

\vskip 0.5cm

\date{\today}

\begin{abstract} We consider the impact of some known extensions of General Relativity in observables that will be available with the
Galileo positioning systems, and draw conclusions as to the
possibility of measuring them. We specifically address the effects of
the presence of a cosmological constant, a Yukawa-like addition to the
Newtonian potential, and the existence of an extra, constant
acceleration. We also consider the phenomenological impact of a broad
class of metric theories, which can be classified through the
parameterised Post-Newtonian formalism.

\vskip 0.5cm

\end{abstract}

\maketitle


\section{Introduction}

The Galileo positioning system poses a great opportunity, not only for
the improvement and development of new applications in navigation monitoring and
related topics, but also possibly for fundamental research in physics. Indeed,
together with the already deployed american and russian counterparts,
the Global Positioning System (GPS) and Glonass, satellite navigation
may be considered the first practical application where relativistic
effects are taken into account, not from an experimental point of
view, but as a regular engineering constraint on the overall design
requirements. Indeed, effects arising from special and General
Relativity (GR) -- gravitational blueshift, time dilation and Sagnac
effect -- may account to as much as $\sim 40~\mu s/day$, which is many
orders of magnitude above the accuracy of the onboard clock deployed
in these systems. Moreover, the gravitational Doppler effect, of the
order of $V_N /c^2 \sim 10^{-10}$ (where $V_N = GM_E /R_E$ is the
Newtonian potential, $G$ is Newton's constant, $M_E \approx 6.0 \times
10^{24}~kg$ is the Earth's mass, $R_E \approx 6.4 \times 10^6~m$ is
its radius and $c$ is the speed of light) falls within the $10^{-12}$
frequency accuracy of current space-certified clocks, and must
therefore be taken into account: in GPS, this is done by imposing an
offset in the onboard clock frequency, while in Galileo this
correction should be corrected by the receiver. For further details,
the reader is directed to Refs. \cite{Ashby,Pascual,Rovelli,Bahder} and references
within.

This said, it is not clear as to what extent the accuracy
of the Galileo positioning system may be improved -- which is designed to offer pinpoint localisation within
an error margin of $1~m$, against the $10~m$ margin of previous the
GPS system -- so to provide clues to the nature of models beyond the current
GR scenario. In this study, we aim at establishing some bounds on the
detectability of commonly considered extensions to GR \cite{review}. This paper is
organised as follows: firstly, we assess the main relativistic effects
that are present in the Galileo system. We proceed and consider the
possibility of measuring a variety of extensions of GR and conclusions are then drawn.

\section{Main relativistic effects}

\subsection{Frame of reference}

Assuming that time-dependent effects are of cosmological origin, and
hence of order $H_0^{-1}$, where $H_0$ is Hubble's constant, one may
discard these as too small within the timeframe of interest; hence, one
assumes a static, spherically symmetric scenario, posited by the
standard Scharzschild metric. In isotropic form, this is given by the
line element

\beq ds^2 = -\left(1 + {2V\over c^2} \right)(c~dt)^2 + {1 \over 1+ {2V
\over c^2} } dV \cong -\left(1 + {2V\over c^2} \right)(c~dt)^2 +
\left(1- {2V \over c^2} \right) dV ~~,\eeq

\noindent where $dV = dr^2 + d\Om^2$ is the volume element, and $V$ is
the gravitational potential. In the standard GR scenario, the latter
coincides with the Newtonian potential $V = V_N = -GM_E /r (1 +
\Sigma^n_{i=1}J_n)$, where the $J_n$ multipoles account for the effect
of geographic perturbations and density profiles.

However, one must introduce the rotation of the Earth with respect to
this fixed-axis reference frame, with angular velocity $\om= 7.29
\times 10^{-5}~rad/s$; by doing a coordinate shift $t' =t$, $r' = r$,
$\th'=\th$ and $\phi' = \phi - \om t'$, one gets the Langevin metric,
given by the line element

\beq ds^2 = -\left[1 + {2V\over c^2} - \left({\om r \sin \th \over
c}\right)^2 \right](c~dt)^2 + 2 \om r^2 \sin^2\th d\phi dt+\left(1 +
{2V \over c^2}\right) dV~~,\eeq

\noindent where, for simplicity, primes were dropped. Asides from a
non-diagonal element, one obtains an addition to the gravitational
potential, which could be viewed as a centrifugal contribution due to
the rotation of the reference frame. One can then define an effective
potential $\Phi = 2V - (\om r sin \th)^2$; the parameterization of the
Earth's geoid is obtained by taking the multipole expansion of $V$ up
to the desired order and finding the equipotential lines $\Phi
=\Phi_0$ (the latter being the value of $\Phi$ at the equator), and
solving for $r(\th,\phi)$.

In the above line elements, the coordinate time coincides with the
proper time of an observer at infinity. However, since one wishes to
evaluate the ground to orbit clock synchronisation, it is advantageous
to rewrite the metric in terms of a rescaled time coordinate, which
coincides with the proper time of clocks at rest on the surface of the
Earth; this is best implemented by resorting to the above-mentioned
geoid, since its definition as an equipotential surface $\Phi =
\Phi_0$ indicates that all clocks at rest in it beat at the same rate;
hence, rescaling the time coordinate according to $t \rightarrow (1
+\Phi_0 / c^2) t$, one gets the metric given by the line element

\beq ds^2 = - \left[ 1 + {2 (\Phi-\Phi_0) \over c^2} \right] (c ~dt)^2
+ 2 \om r^2 sin^2\th d\phi dt + \left(1 - {2V\over c^2}
\right)d\Om~~. \eeq

Finally, if one reassumes a non-rotating frame, the metric is given by
the line element

\beq ds^2 = - \left[ 1 + {2 (V-\Phi_0) \over c^2} \right] (c ~dt)^2 +
\left(1 - {2V\over c^2} \right)d\Om~~. \eeq

\subsection{Constant and periodic clock deviation}

One may now consider the difference between the time elapsed on the
ground and the satellite clock; keeping only terms of order $c^{-2}$,
one finds that the proper time increment on the moving clock is
approximately given by

\beq d\tau = ds / c = \left( 1 + {V-\Phi_0 \over c^2} -{v^2 \over 2
c^2} \right) dt~~.\eeq

\noindent Considering an elliptic orbit with semi-major axis $a$, and
taking $V = V_N \approx GM_E / r$, this may be recast into the form
\cite{Ashby}

\beq d\tau = ds / c = \left[ 1 + {3GM_E \over 2 a c^2} + {\Phi_0 \over
c^2} - {2GM_E \over c^2} \left( {1 \over a } - {1 \over r}\right)
\right] dt~~.\eeq

\noindent The first constant rate correction terms in the above amount
to \beq {3GM_E \over 2 a c^2} + {\Phi_0 \over c^2} = -4.7454 \times
10^{-10}~~,\eeq

\noindent for the Galileo system, and $-4.4647 \times 10^{-10}$, for
the GPS system; this indicates that the orbiting clock is beating
faster, by about $41~\mu s /day$, for the Galileo system, and
$39~\mu s /day$, for the GPS system . For this reason, the GPS system
has a built in frequency offset of this magnitude, while the increased
computational capabilities made available to current and future
receivers of the Galileo system leave this correction to the user. The
residual periodic corrections, proportional to $1/r - 1/a$, have an
amplitude of order $49~ns/day$, for the Galileo system, and
$46~ns/day$, for the GPS system.

\subsection{Shapiro time delay and the Sagnac effect}

The so-called Shapiro time delay, a second order relativistic effect
due to the signal propagation is given by \cite{Ashby}

\beq \De t_{delay} = {\Phi_0 l \over c^3} + {2GM_E \over c^3}~ln
\left(1 + {l \over R_E} \right)~~,\eeq

\noindent where we have integrated over a straight line path of
(proper) length $l$. Evaluating this delay, one concludes that this
effect amounts to $6.67 \times 10^{-11}~s$.

Also, one must consider the so-called Sagnac effect, which arises from
the difference between the gravitational potential $V$ and the
effective potential $\Phi$, when proceeding from a non-rotational to a
rotational frame. Hence, one gets the additional time delay

\beq \De t_{Sagnac} = { \om \over c^2} \int_{path} r^2 ~d\phi = {2 \om
\over c^2} \int_{path} dA_z ~~, \eeq

\noindent where $dA_z$ is the orto-equatorial projection of the area
element swept by a vector from the rotation axis to the satellite. For
the Galileo system, this yields a maximum value of $153~ns$ while, for
the GPS system, one gets $133~ns$.

One concludes this section by recalling the main effects affecting the
considered global positioning systems: a frequency shift of order
$ 10^{-10}$ and a propagation time delay (Shapiro plus Sagnac effect)
of the order $10^{-7}~s$. In what follows, one shall compute the
additional frequency shift and propagation time delay induced by
common proposals for extensions of GR, and compare the results with
the above quantities, plus the frequency accuracy of $10^{-12}$ and
the time accuracy of Galileo, of order $10^{-9}~s$, which corresponds to a
optimistic spatial accuracy of $30~cm$.

\subsection{Post-Newtonian effects}

We address now the issue of measuring Post-Newtonian effects with the Galileo positioning system. As the moniker indicates, these are effects below the Newtonian order, that is, $GM_E/R_Ec^2 \approx 10^{-10}$. A general formalism exists to describe lower-order effects induced by extensions to GR and alternate theories that resort to a metric approach of gravity; indeed, any such theory may be analysed locally and compared with the so-called Parameterised Post-Newtonian (PPN) metric \cite{Will,Klioner}, given by the line element

\beq ds^2 = - \left[ 1 - {2V \over c^2} + 2 \be \left({ V \over c^2}\right)^2 \right]~(c~dt)^2 + \left( 1 - 2 \ga {V \over c^2} \right)~dV ~~.\eeq

\noindent In the above, the parameter $\be$ measures the non-linearity of the superposition law for gravity, while $\ga$ indicates the space curvature produced per unit mass. For clarity, we consider only a simplified version of the full PPN metric; the latter encompasses ten PPN parameters, characterising the underlying fundamental theory, and may be related to violation of momentum conservation, existence of a privileged reference frame, amongst others. GR is characterised by $\be=\ga=1$, while all remaining parameters vanish. For that reason, the quantities $\be -1$ and $\ga - 1$ measure the deviation from the predictions of the currently accepted theory. Experimentally, it is found that $|\be -1 | \leq 2-3 \times 10^{-4}$ (Nordtvedt effect) and $\ga -1 = (2.1 \pm 2.3) \times 10^{-5}$ (Cassini radiometry).

Unfortunately, it is clear from the above equation that Post-Newtonian effects arise only at an order $\sim 10^{-20}$, undetectable by the accuracy of the GPS and Galileo systems. 

\section{Detection of the cosmological constant}

Latest observations indicate that the Universe is experiencing an accelerated expansion, which may be characterised by the presence of a cosmological constant $\La \sim 10^{-35}~s^{-2}$, acting as a negative-pressure fluid (see {\it e.g.} \cite{Lambda} and references therein). By matching the outer Friedmann-Robertson-Walker metric with a static, symmetric solution given by Birkhoff's theorem, we may derive the Schwarschild-de Sitter metric, given by the line element (in anisotropic form) \cite{SdS},

\beq ds^2 = -\left(1-{2V_N \over c^2 } - {\La r^2\over 3c^2}\right) (c~dt)^2 + {1 \over 1-{2V_N \over c^2 } - {\La r^2\over 3}}dr^2 + d\Om~~.\eeq

\noindent This indicates that the cosmological constant induces an additional term to the potential, of the form $V_\La = -\La r^2/6$; since its expected effect is assumed to be small, one may neglect the issue of performing a coordinate change to an isotropic, co-rotating frame of reference, as well as the identification of proper time with clocks at rest on the surface of the geoid (however, notice that the identification of proper time as that measured by a clock at rest at infinity breaks down, due to the Schwarschild ``bubble'' breaking down at a distance $r$ given by the condition $V_n= V_\La$).

The frequency shift of a signal emitted at a distance from the origin $r=R_E+h$ (for the Galileo system, $h = 17.2 \times 10^6~m$) and received at a distance $r=R_E$ is given by 

\beq \left({f_{Earth} \over f_{Sat} }\right) = \sqrt{g_{00~Earth} \over g_{00~Sat}} = \sqrt{1 - 2V(R_E)/c^2 \over 1-2V(R_E+h)/c^2} \simeq {V(R_E)-V(R_E+h) \over c^2}~~. \eeq

\noindent Hence, one may compute the additional frequency shift induced by this extra potential contribution, through

\beq \left({f_{Earth} \over f_{Sat} }\right)_\La \simeq {V_\La(R_E) - V_\La(R_E+h) \over c^2} = {\La \over 6c^2}h(2R_E+h) \sim 10^{-38}~~, \eeq

\noindent which clearly falls bellow the accuracy $\ep_{f_r} = 10^{-12}$ of the Galileo constellation.

Likewise, the propagational time delay is given by

\beq \De t_{delay}={1 \over c} \int_{R_E}^{R_E+h} V(r)~dr~~. \eeq

\noindent Hence, the cosmological constant induces a further delay of

\beq \De t_\La = {1 \over c}  \int_{R_E}^{R_E+h} {\La r^2 \over 6 c^2}~dr = {\La \over 18c^3} h \left[ (3R_E(R_E+h) + h^2\right] \sim 10^{-40}~s~~, \eeq

\noindent also many orders of magnitude below the time resolution of $10^{-9}~s$. Therefore, one concludes that the cosmological constant is completely undetectable by the Galileo system.

\section{Detection of anomalous, constant acceleration}

An anomalous constant acceleration could model first-order effects arising from some fundamental theory of gravitation which expands upon GR, or indicate some threshold between known dynamics and yet undetected, exotic physics. One examples stems from the so-called Modified Newtonian Dynamics (MOND) model \cite{Milgrom, Bekenstein,DM}, which attempts to account for the missing matter problem indicated by galactic rotation curves without the need for dark matter, by featuring a departure from the classical Poisson equation at a characteristic acceleration scale of $10^{-10}~m/s^2$. Also, although yet unmodelled or theoretically unaccounted for, an anomalous, sunbound, constant acceleration $a=(8.74 \pm 1.33) \times 10^{-10}~m/s^2$ has been reported to affect the Pioneer 10/11 probes \cite{Slava,Paramos,Reynaud}.

An anomalous, constant acceleration $a$ may be phenomenologically modelled by a potential $V_a = ar$; following the procedure depicted in the previous section, the following frequency shift is obtained

\beq \left({f_{Earth} \over f_{Sat} }\right)_a \simeq {V_a(R_E) - V_a(R_E+h) \over c^2} = {ah \over c^2}~~. \eeq

\noindent Comparing with the frequency accuracy $\ep_{f_r} = 10^{-12}$, one finds that only accelerations $ a \geq c^2 \ep_{f_r} / h \sim 10^{-3}~m/s^2$ may be detected.

The propagational time delay due to this extra potential addition is given by

\beq \De t_a = {1 \over c}  \int_{R_E}^{R_E+h} {ar\over c^2}~dr = {a \over 2 c^3}h(2R_E+h)~~, \eeq

\noindent and comparison with a time accuracy of $10^{-9}~s$ yields the condition for detectability $a \gtrsim 100~m/s^2 $. Therefore, one concludes that accelerations of the order $10^{-10} - 10^{-9}~m/s^2$ are beyond the observable reach of the Galileo system; conversely, detectability of a constant acceleration of the order of $10^{-10}~m/s^2$ would require an improvement of $7$ orders of magnitude in frequency accuracy (to $\ep_{f_r} \sim 10^{-19}$) and $12$ orders of magnitude in time resolution (to $10^{-21} ~s$).

\section{Detection of Yukawa potential}

A common phenomenological approach to extensions of GR lies in assuming that the potential has a modified Yukawa form,

\beq V(r) = -{G_\infty M_E \over r} \left(1 + \al e^{- r /\la} \right)~~, \eeq

\noindent where $\al$ is the strength of the perturbation, $\la$ its characteristic range, and $G_\infty$ the gravitational coupling for $r \rightarrow \infty$; the latter may be regarded as a redefinition of Newton's constant $ G $, through $G = G_\infty (1+\al)$. This potential may be separated into a Newtonian-like potential and an extra potential $V_Y = -(\al G M_E /(1+\al)r)e^{-r/\la)}$. The Yukawa contribution may arise from scalar/tensor field models, where the range is related to the mass $m$ of the scalar field, $\la \propto m^{-1}$ \cite{review}.

\begin{figure} 

\epsfysize=8cm \epsffile{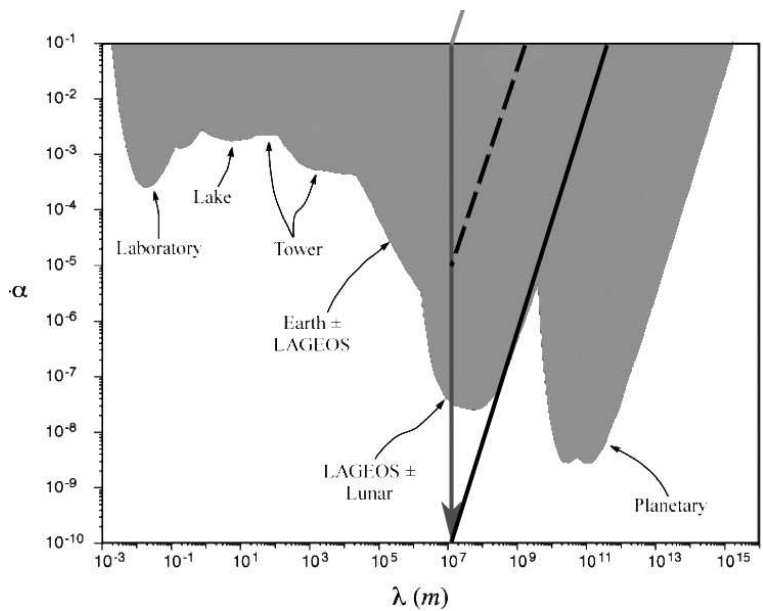}
\caption{Exclusion plot for the Yukawa strength $\al$ and range $\la$, and superimposed limits obtained for varying frequency accuracy $\ep_{f_r}$: $10^{-10}$ (grey, full), $10^{-12}$ (black dash) and $10^{-19}$ (black full).}
\label{yukawa}

\end{figure}

Tight experimental constraints are available, stemming from several sources and regimes, as may be seen in Fig. \ref{yukawa}. Clearly, two yet unexplored avenues remain: the sub-millimeter regime, $\la < 10^{-3}~m$ \cite{Yukawa}, and an astronomical regime, $\la > 10^{15}~m \approx 0.1~ly$.

Following the previous steps, one first obtains the extra frequency shift

\beq \left({f_{Earth} \over f_{Sat}} \right)_Y = {V_Y(R_E) - V_Y(R_E+h) \over c^2} = {GM_E \over c^2 R_E}\left({\al \over 1+\al}\right) e^{-R_E/\la} \left(e^{-h/\la}{R_E \over R_E + h} - 1 \right)~~.\eeq

The additional time delay is given by

\beq \De t_Y = {1 \over c} \int_{R_E}^{R_E+h} {GM_E \over c^2 r }\left({\al \over 1+\al}\right) e^{-r/\la}~dr~~. \eeq
The above expressions may be considerably shortened if it is assumed that this additional ``fifth-force'' is a long-range, $\la \gg r$, or short-range interaction, $\la \ll r$.

\subsection{Short-range fifth force}

If the range of the Yukawa interaction is short-ranged, $\la \ll h,~R_E$, one obtains

\beq \left({f_{Earth} \over f_{Sat}} \right)_Y \simeq -{GM_E \over c^2 R_E}\left({\al \over 1+ \al}\right) e^{-R_E/\la}~~.\eeq

\noindent If this effect is undetectable within the frequency accuracy $\ep_{f_r}$, one obtains the constraint for small $\al$

\beq \al \lesssim \left[{GM_E \over c^2 R_E} \right]^{-1} e^{R_E/\la} \ep_{f_r} \approx 1.4 \times 10^{-3} e^{R_E/\la} \gg 1~~,\eeq

\noindent which yields no new insight into the yet unexplored sub-millimetric regime, as depicted in Fig. \ref{yukawa}.

Likewise, the additional propagation time delay is given by

\beq \De t_Y = - {GM_E \al \over c^3} ~ln\left(1+ {h \over R_E}\right)~~, \eeq

\noindent so that comparison with the time accuracy of $\De t= 10^{-9}$ yields, for $\al \ll 1$

\beq \al \leq \left[ {GM_E \over c^3} ~ln\left(1+ {h \over R_E}\right)\right]^{-1} \De t \approx 50~~. \eeq

Hence, one concludes that the short-range regime of a hypothetical Yukawa fifth force cannot be probed by the Galileo system.

\subsection{Long-range fifth force}

If one follows the inverse assumption of the previous subsection, and assumes a long range fifth force, $\la \gg h, R_E$, the exponential terms may be expanded to first order in $r/\la$; as a result, the induced propagation time delay becomes

\beq \De t_Y \simeq -{GM_E \al \over c^3} {h \over \la}~~.\eeq

\noindent If the effect is undetected at a level of accuracy $\De t \sim 10^{-9}~s$, one obtains, for small $\al$

\beq |\al | < {c^3 \De t \over GM_E}{\la \over h} \approx 4 \times 10^{-6}\left( {\la \over 1~m} \right)~~. \eeq

\noindent For a lower bound of $\la \approx 10^8~m$ (only one order of magnitude above $R_E, h$), we obtain the incompatible result $\al < 400$.

Regarding the additional frequency shift, one finds

\beq \left({f_{Earth} \over f_{Sat}} \right)_Y \simeq {GM_E \al h\over 2 c^2 \la^2}~~,\eeq

\noindent so that comparison with the frequency accuracy level of $\ep_{f_r} \sim 10^{-12}$ yields, for $\al \ll 1$

\beq \al < \left({GM_E \over c^2} \right)^{-1}\left({2 \la^2 \over h} \right) \ep_{f_r} \approx 10^{-5}\ep_{f_r} \left({\la \over 1~m}\right)^2 ~~,\eeq 

\noindent or, equivalently, a quite interesting bound

\beq log ~\al < -5 + log ~\ep_{f_r}+ 2 ~log \left( {\la \over 1~m } \right) ~~. \eeq

One may plot the different constraints obtained by varying the frequency accuracy $\ep_{f_r}$, as seen in Fig. \ref{yukawa}; this shows that, at the current level, no new bounds are produced. Also, it shows that, at a level $\ep_{f_r} \sim 10^{-19}$, the region below the ``trough'' at $\la \sim 10^8~m$ (corresponding to $\al < 10^{-8}$) could be investigated.

\section{Conclusions}

In this work, we have addressed the possibility of detecting signals of new physics through the use of the Galileo positioning system. This application could be valuable, as any unexpected new phenomenology could provide further insight into what lies beyond General Relativity. We have specifically looked at the propagation time delay and frequency shift induced by three different models, namely a potential related to the presence of the cosmological constant, the influence of an anomalous, constant acceleration, and the addition of a Yukawa-like fifth force. We also briefly discussed the (im)possibility of measuring post-Newtonian effects with the Galileo system.

Unfortunately, our conclusions indicate that the available observables are not suitable for the intended purpose; indeed, while these render the detection of the cosmological constant unattainable, they also indicate that the current accuracy is many orders of magnitude above that needed to probe interesting regimes of anomalous constant acceleration ($a \sim 10^{-10} - 10^{-9} ~m/s^2$) or Yukawa range $\la > 10^{8}~m$ and strength $\al < 10^{-8}$. Indeed, a frequency accuracy of $10^{-19}$, near the ``quantum'' regime, is required to further probe the desired scales. Although this is clearly beyond the grasp of any foreseeable global positioning systems, and yet unavailable in space certified clocks, such precision might be attainable in the future.

Finally, we remark that, although it was not the purpose of this study, the Galileo positioning system could be paramount in improving the bound on violation of the Local Positioning Invariance (LPI) principle \cite{review}; this tenant, one of the fundamental pillars of General Relativity, postulates that clock rates are independent of their spacetime positions. Experimental constraints on allowed relative frequency deviations indicate that this invariance holds down to a level of $2.1 \times 10^{-5}$ \cite{LPI}. Endowing one or more elements of the Galileo constellation with higher precision clocks and allowing for sufficiently stable communication with stations on Earth, possibly through a microwave link, could yield an improvement of up to two orders of magnitude on the LPI. Another alternative could involve installing cornercubes on the surface of one or more elements of the Galileo system, so to allow for accurate laser ranging. It is tempting to call this subset of the Galileo constellation {\it Siderius Nuncius}, the Celestial Messenger, given its potential in helping to sort out the mysteries of the Cosmos.

\begin{acknowledgments}

The work of J.P. is sponsored by the Funda\c{c}\~ao para a Ci\^encia e Tecnologia (FCT) under the grant $BPD~23287/2005$. O.B. acknowledges the partial support of the FCT project $PDCTE/FNU/50415/2003$.

\end{acknowledgments}


\begin{thebibliography}{99}


\bibitem{Ashby}N. Ashby, {\it Liv. Rev. Rel.} {\bf 6} (2003) 1.

\bibitem{Pascual}J. Pascual-Sanchez, gr-qc/0507121.

\bibitem{Rovelli}C. Rovelli, \PR {\bf D 65} (2002) 044017.

\bibitem{Bahder}T. Bahder, \PR {\bf D 68} (2003) 063005.

\bibitem{review}O. Bertolami, J. P\' aramos, S. Turyshev, gr-qc/0602016.

\bibitem{Will}C. Will, {\it ``Theory and Experiment in Gravitational Physics''}, Cambridge U. P. (1993).

\bibitem{Klioner}S. Klioner and M. Soffel, \PR {\bf D 62} (2000) 024019.

\bibitem{Lambda}M. Bento, O. Bertolami, N. Santos, A. Sen, \PR {\bf D 71} (2005) 063501.

\bibitem{SdS}S. Hawking and G. Gibbons, \PR {\bf D 15} (1977) 2738.

\bibitem{Milgrom}M. Milgrom, \AJ {\bf 270} (1983) 365.

\bibitem{Bekenstein}J. Bekenstein, \PR {\bf D70} (2004) 083509.

\bibitem{DM}See {\it e.g.} O. Bertolami and J. P\'aramos, gr-qc/0611025, for a critical assessment, and O. Bertolami, C. B\"ohmer, T. Harko and F. Lobo, \PR {\bf D 75} (2007) 104016, for an alternative view of the MOND approach.
 
\bibitem{Slava}J. Anderson et al., \PRL {\bf 81} (1998) 2858; \PR {\bf D 65} (2002) 082004.

\bibitem{Paramos}See {\it e.g.} O. Bertolami and J. P\'aramos, \CQG {\bf 21} (2004) 3309, for a solution involving a scalar field with a suitable potential, and for a list of other possible solutions.

\bibitem{Reynaud}Other solutions include M. Jaekel and S. Reynaud, \MPL {\bf A 20} (2005) 1047; J. Brownstein and J. Moffat, \CQG {\bf 23} (2006) 3427.

\bibitem{Yukawa}See {\it e.g.} E. Adelberger, B. Heckel and A. Nelson, {\it Ann.Rev.Nucl.Part.Sci.} {\bf 53} (2003) 77, for a discussion and the latest experimental bound on non-Newtonian forces at sub-millimeter range, $\la < 0.3 ~mm$ for $ \al = 16/3$.

\bibitem{LPI}A. Bauch and S. Weyers, \PR {\bf D 65} (2002) 081101.

\end{thebibliography}
\end{document}